\pdfoutput=1
\documentclass{article}
\usepackage{spconf,amsmath,graphicx,hyperref}
\usepackage{amssymb}
\usepackage{arydshln}
\newcommand{\seq}[3]{[#1_{#2}]_{#2=1}^{#3}}

\def\m{{\mathbf m}}

\def\h{{\mathbf h}}

\def\bmH{{\mathbf H}}

\def\enc{\mathsf{e}}

\def\calL{{\cal L}}
\def\hs{{\h_{\mathsf{s}}}}

\def\para{\vspace{.6mm}\noindent}
\title{
REVISITING FRAME-WISE SALIENCY FOR AUDIO MOMENT RETRIEVAL
}
\name{Tatsuya Komatsu, Hokuto Munakata}
\address{LY Corporation, Tokyo, Japan}

\begin{document}
\ninept
\maketitle

\begin{abstract}
This paper revisits frame-wise saliency for audio moment retrieval (AMR).
We show that the frame-wise saliency sequence, conventionally used only as an auxiliary output in DETR-based AMR models, can itself serve as an effective source of moment predictions.
We convert the saliency sequence into ranked moments using a simple SED-inspired segmentation rule with no learned parameters, enabling moment retrieval directly from frame-wise temporal information.
On the CASTELLA dataset, saliency-based prediction consistently outperforms decoder-based prediction from the same model across all 18 runs of QD-DETR and CG-DETR.
For QD-DETR, simply replacing the inference output improves R1@0.7 from 21.0 to 36.1. The advantage remains 7–15 points when the two outputs are evaluated at their independently selected best epochs.
The same tendency extends to TaskWeave and UVCOM, whereas TR-DETR shows the opposite behavior, suggesting that how saliency construction may matter.
The performance gap is especially pronounced for short moments: for queries whose annotated moments average at most 2 s, R1@0.7 improves from 6.3 to 27.8 with QD-DETR.
Decoder supervision nevertheless benefits saliency-based prediction, indicating that its role during training differs from the utility of its inference output.
\end{abstract}

\section{Introduction}
\label{sec:intro}
Given a natural-language query and a long audio recording, audio moment retrieval (AMR) returns the temporal segments, or moments, described by the query~\cite{munakata2025amr}.
Unlike sound event detection (SED), which detects instances of predefined sound event classes~\cite{mesaros2021sed}, AMR supports free-form text queries.
Such queries can describe multiple sounds and their relationships, as in ``a rooster crows and pigeons coo'' or ``a car passes by while children are playing in the distance.''
Moreover, multiple distinct intervals within a recording may match the same query.
These characteristics require capturing audio--query relevance over time and using this time-varying relevance to identify the temporal boundaries of individual moments.

Many AMR models inherit the DETR-style encoder--decoder architecture developed for video moment retrieval.
In the DETR-style architecture, the encoder produces query-conditioned frame-wise representations, while the decoder predicts a fixed set of temporal segments.
Moment-DETR~\cite{lei2021momentdetr} adapted the DETR set-prediction framework~\cite{carion2020detr} to video moment retrieval, using one-to-one matching between predicted and ground-truth moments.
QD-DETR~\cite{moon2023qddetr} strengthened the text conditioning of the frame representations, and CG-DETR~\cite{moon2023cgdetr} further refined these representations using clip--word correlations.
AM-DETR~\cite{munakata2025amr} subsequently adapted the QD-DETR framework to audio, and variants of this DETR-based formulation were further explored in the DCASE 2026 AMR task~\cite{munakata2026dcase}.
These models also employ a saliency head that assigns a relevance score to each frame, providing auxiliary supervision for the frame-wise representations~\cite{lei2021momentdetr}.
At inference time, however, temporal segments are predicted only by the decoder, and these frame-wise saliency scores are not directly used to determine the temporal boundaries.

In SED, by contrast, frame-wise event probabilities are directly converted into event onsets and offsets through thresholding and temporal smoothing~\cite{mesaros2021sed, cakir2017crnn,miyazaki2020conformer}.
Both approaches produce temporal segments, but SED derives boundaries directly from frame-wise predictions without a learned moment decoder.
This suggests an analogous approach to AMR: directly converting text-conditioned frame-wise saliency scores into moments rather than relying on the decoder to produce the final moments.
It remains unclear whether the decoder predicts moments more accurately than a rule that derives them directly from these frame-wise scores.

In this paper, we investigate this question by segmenting the frame-wise saliency sequences of trained DETR-based AMR models.
A fixed, SED-inspired segmentation rule with no learned parameters converts these saliency sequences into saliency-based predictions, which are then compared with the decoder-based predictions from the same trained models.
Experiments are conducted on CASTELLA~\cite{munakata2025castella}, a human-annotated AMR benchmark, across several DETR-based AMR models.
We analyze how the performance gap between the two approaches varies with the characteristics of the target moments and with how the saliency sequence is constructed.
The segmentation rule is further compared with simpler alternatives applied to the same saliency sequences.
The frame-wise representations are shared by both the saliency head and the moment decoder.
We therefore also examine how saliency and decoder supervision affect saliency-based prediction.

\section{Related Work}
\label{sec:related}
\noindent\textbf{Temporal localization.}
DETR-based video and audio moment retrieval models typically predict temporal segments from a fixed set of decoder queries~\cite{munakata2025amr,lei2021momentdetr,moon2023qddetr,moon2023cgdetr,lee2024bamdetr,jang2023eatr}.
In several such models, frame-wise saliency is additionally predicted as an auxiliary output~\cite{lei2021momentdetr,moon2023qddetr}.
Some studies make greater use of frame-wise information: SG-DETR~\cite{gordeev2024sgdetr} uses saliency to guide decoder attention, while TR-DETR~\cite{sun2024trdetr} and TaskWeave~\cite{yang2024taskweave} exchange information between highlight detection and moment retrieval.
Nevertheless, these models still obtain the final moments from the decoder.

Frame-wise temporal localization provides a different formulation.
Sound event detection derives event boundaries from frame-wise probabilities~\cite{mesaros2021sed,cakir2017crnn,miyazaki2020conformer}, and text-to-audio grounding localizes described sounds frame by frame~\cite{xu2021tag}.
In video, span-based and anchor-free models likewise make temporal predictions at each time step~\cite{zhang2022actionformer,zhang2020vslnet}.
UniVTG~\cite{lin2023univtg}, for example, applies frame-wise foreground and boundary prediction to video moment retrieval without a DETR decoder.
Together, these works show that temporal localization can be performed directly from frame-wise predictions, without treating them solely as auxiliary signals for decoder-based localization.
Our approach differs in that we do not train a separate frame-wise localization architecture; instead, we segment the saliency sequence already produced by a trained DETR-based model.

\noindent\textbf{Audio moment retrieval.}
AMR was introduced with AM-DETR~\cite{munakata2025amr} and subsequently featured as Task 6 of the DCASE 2026 Challenge~\cite{munakata2026dcase}.
Participants explored stronger audio--text encoders~\cite{niizumi2024m2dclap,kibata2026_t6,wu2023laionclap,kang2026_t6,wu2025flam}, improved training and ensembling~\cite{usui2026_t6,calvet2026_t6,sugawara2026_t6}, and adopted several localization architectures from video moment retrieval~\cite{moon2023cgdetr,lee2024bamdetr,kibata2026_t6,sugawara2026_t6,ogawa2026_t6,xiao2024uvcom}.
Some systems instead derived moments directly from pretrained audio--text similarity without a DETR-based network~\cite{huck2026_t6,kret2026_t6}.
Concurrent work by Choi et al.~\cite{choi2026segmental} replaces DETR decoding at inference with a learned segmental model based on frame-wise scores.
In contrast, we derive moments from the saliency already produced by trained DETR-based AMR models and directly compare them with decoder predictions from the same models, without training a replacement moment prediction model.

\section{DETR-based audio moment retrieval}
\label{sec:detr}

In QD-DETR~\cite{moon2023qddetr} and CG-DETR~\cite{moon2023cgdetr}, our primary models, moment prediction and frame-wise saliency estimation share the encoded audio representations.
Given projected audio and text features, the encoder combines cross-modal attention with temporal self-attention to produce text-conditioned frame representations
$\bmH_\enc=\seq{\h^\enc}{j}{L_a}\in\mathbb{R}^{L_a\times d}$
and a global representation $\hs\in\mathbb{R}^{d}$, where $L_a$ is the number of audio frames and $d$ is the hidden dimension.
A frame denotes one temporal feature step (1~s in our experiments).
Here, $\bmH_\enc$ denotes the final frame representations supplied to both the decoder and the saliency head; for CG-DETR, this includes adding $\hs$ to each frame.

The decoder starts from $N_a$ learned moment anchors (the decoder queries), each specifying an initial center and duration, with associated content vectors initialized to zero.
It attends to $\bmH_\enc$ and refines the anchors and their content vectors across its layers.
The resulting candidates are
$\{(\hat{\m}_i,s_i)\}_{i=1}^{N_a}$,
where $\hat{\m}_i=(\hat{t}_i,\hat{w}_i)$ contains the predicted center and duration, normalized by the recording length, and $s_i\in[0,1]$ is the foreground confidence.
At inference, these candidates are ranked by $s_i$ to produce the decoder-based predictions.

The saliency head instead assigns a score to each frame directly from the encoder representations, without using decoder predictions.
Following QD-DETR and CG-DETR~\cite{moon2023qddetr,moon2023cgdetr}, it computes a scaled dot product between projected frame and global representations.
At inference, our segmentation procedure applies a fixed sigmoid transform to the raw scores $r_j$, yielding bounded scores $p_j$:
\begin{equation}
r_j =
\frac{f_{\mathsf{f}}(\h_j^\enc)^\top
      f_{\mathsf{g}}(\hs)}{\sqrt{d}},
\qquad
p_j = \sigma(r_j)\in(0,1),
\label{eq:saliency}
\end{equation}
where $f_{\mathsf{f}}$ and $f_{\mathsf{g}}$ are learned affine projections from $\mathbb{R}^{d}$ to $\mathbb{R}^{d}$, and $\sigma$ is the logistic sigmoid.
Standard inference does not use this saliency sequence to determine moment boundaries.
We instead convert it into ranked moments using the segmentation rule in Sec.~\ref{ssec:segmentation}.

The models are jointly trained with
\begin{equation}
\calL =
\calL_{\mathsf{dec}}
+ \lambda\,\calL_{\mathsf{sal}}
+ \calL_{\mathsf{other}},
\label{eq:loss}
\end{equation}
where $\lambda>0$ controls the weight of saliency supervision.
The decoder objective $\calL_{\mathsf{dec}}$ combines span-regression and generalized-IoU losses for candidates matched to annotated moments with foreground/background classification for all candidates.
Matches are determined by one-to-one bipartite matching, with unmatched candidates labeled as background.
The saliency objective $\calL_{\mathsf{sal}}$ groups the models' frame-wise supervision terms with their original internal weights, and $\lambda$ scales this objective as a whole.
The remaining model-specific auxiliary losses, including CG-DETR's alignment and distillation terms, are grouped in $\calL_{\mathsf{other}}$; $\calL_{\mathsf{other}}=0$ for QD-DETR.
We retain the constituent loss terms and their internal weights except in the training-objective experiments of Sec.~\ref{ssec:exp_training}.
Our primary comparison thus holds the trained model fixed while evaluating two outputs that differ in both prediction mechanism and supervision.

\section{Segmenting the saliency sequence into moments}
\label{sec:method}

\subsection{Peak-relative segmentation}
\label{ssec:segmentation}
Given the frame-wise saliency scores $\{p_j\}_{j=1}^{L_a}$ in Eq.~(\ref{eq:saliency}), we convert the sequence into ranked temporal moments.
Instead of using a single recording-level threshold, we segment around individual peaks using a peak-relative threshold.
This allows weaker peaks to form moments even when stronger peaks occur elsewhere in the recording.
\begin{itemize}
\item \textbf{Detect peaks.} Each local maximum $k$ forms a candidate moment, with the first and last frames also eligible as peaks.
\item \textbf{Segment.} For each peak, we set $\theta_k=\alpha p_k=\alpha\sigma(r_k)$ and extend the candidate in both temporal directions until $g+1$ consecutive frames fall below $\theta_k$, excluding this final run from the candidate.
\item \textbf{Rank and remove overlaps.} Candidates are scored and ranked by the mean of $p_j$ over all frames in each candidate. In descending score order, a candidate is removed if its temporal IoU with any retained higher-ranked candidate exceeds $\tau$.
\end{itemize}
We refer to the resulting moments as \emph{saliency-based predictions}, in contrast to the \emph{decoder-based predictions} produced by the DETR decoder.
The procedure has three hyperparameters: $\alpha\in(0,1)$ controls the peak-relative threshold, $g\in\{0,1,\ldots\}$ controls the tolerance for short below-threshold gaps, and $\tau\in(0,1)$ controls overlap removal.
The procedure has no learned parameters and is applied unchanged to all models.

\subsection{Training objective for saliency-based prediction}
\label{ssec:training}
We examine the effects of the training objective on saliency-based prediction.
We evaluate the saliency-loss weight in Eq.~\eqref{eq:loss} at $\lambda\in\{1,4,8\}$, starting from the inherited $\lambda=1$~\cite{moon2023qddetr,moon2023cgdetr}.
We also remove decoder supervision by setting $\calL_{\mathsf{dec}}=0$, while retaining $\lambda\calL_{\mathsf{sal}}+\calL_{\mathsf{other}}$, to examine how decoder training affects saliency-based prediction.

\begin{table}[t]
\centering
\caption{
Saliency- vs. decoder-based predictions on CASTELLA test (\%, mean $\pm$ SD over three seeds). $\lambda$: saliency-loss weight.
sal./dec.: saliency-/decoder-based predictions.
}
\label{tab:main}
\small
\scalebox{0.8}{
\setlength{\tabcolsep}{3.5pt}
\begin{tabular}{llcccccc}
\hline
 & & \multicolumn{2}{c}{R1@0.5} & \multicolumn{2}{c}{R1@0.7} & \multicolumn{2}{c}{mAP} \\
Model & $\lambda$ & sal. & dec. & sal. & dec. & sal. & dec. \\
\hline
QD-DETR & 1 & 50.1$\pm$1.1 & 40.8$\pm$1.6 & 36.1$\pm$0.8 & 21.0$\pm$1.3 & 30.7$\pm$0.5 & 17.5$\pm$1.0 \\
        & 4 & 50.3$\pm$0.6 & 39.1$\pm$2.5 & 36.1$\pm$0.8 & 20.2$\pm$2.7 & 30.5$\pm$0.2 & 17.4$\pm$1.5 \\
        & 8 & 51.3$\pm$0.5 & 41.9$\pm$1.5 & 37.3$\pm$1.0 & 21.1$\pm$1.3 & \textbf{31.4}$\pm$0.4 & 18.5$\pm$0.5 \\
CG-DETR & 1 & 51.2$\pm$1.3 & 46.3$\pm$2.9 & 36.7$\pm$0.5 & 25.9$\pm$3.7 & 29.7$\pm$0.5 & 21.3$\pm$2.4 \\
        & 4 & 52.0$\pm$1.0 & 48.2$\pm$4.3 & 38.3$\pm$0.9 & 27.4$\pm$5.7 & 30.6$\pm$0.4 & 22.4$\pm$3.3 \\
        & 8 & \textbf{52.7}$\pm$0.8 & \textbf{48.9}$\pm$1.7 & \textbf{38.5}$\pm$0.6 & \textbf{28.4}$\pm$2.2 & \textbf{31.4}$\pm$0.1 & \textbf{23.0}$\pm$1.7 \\
\hline
\end{tabular}
}\vspace{-4mm}
\end{table}
\begin{table}[t]
\centering
\caption{Decoder-based test R1@0.7 (\%, mean over three seeds) at epochs selected by decoder-based validation R1@0.7. Compare with the saliency-selected results in Table~\ref{tab:main}.}
\label{tab:epochs}
\small
\setlength{\tabcolsep}{3pt}
\begin{tabular}{lcc}
\hline
$\lambda$ & QD-DETR & CG-DETR \\
\hline
1 & 21.4 & 29.8 \\
4 & 23.2 & 30.2 \\
8 & 24.3 & 28.2 \\
\hline
\end{tabular}\vspace{-5mm}
\end{table}

\begin{table}[t]
\centering
\caption{
Saliency-based test R1@0.7 (\%) under different segmentation rules, with decoder-based prediction shown for reference (mean over three seeds).
Simple recording-wise thresholding retains a substantial advantage over decoder-based prediction.
}

\label{tab:variants}
\small
\setlength{\tabcolsep}{3pt}
\begin{tabular}{lcccc}
\hline
& \multicolumn{2}{c}{QD-DETR} & \multicolumn{2}{c}{CG-DETR} \\
& $\lambda{=}1$ & $\lambda{=}8$ & $\lambda{=}1$ & $\lambda{=}8$ \\
\hline
Peak-relative threshold~(Ours) & \textbf{36.1} & \textbf{37.3} & \textbf{36.7} & \textbf{38.5} \\
~~- w/o gap tolerance and overlap removal & 30.2 & 32.1 & 31.3 & 33.4 \\
Recording-wise threshold & 32.5 & 33.0 & 32.7 & 35.4 \\
Global absolute threshold & 21.2 & 24.1 & 25.1 & 24.3 \\
Fixed 5~s duration & 6.1 & 6.2 & 6.0 & 6.5 \\\hdashline
Decoder-based prediction & 21.0 & 21.1 & 25.9 & 28.4 \\
\hline
\end{tabular}
\vspace{-5mm}
\end{table}

\begin{table}[t]
\centering
\caption{
Saliency- and decoder-based predictions for three additional DETR-based models on CASTELLA test (\%, mean over three seeds).
Saliency-based prediction is higher for TaskWeave and UVCOM, whereas TR-DETR shows the opposite behavior. sal./dec.: saliency-/decoder-based predictions.
}
\label{tab:families}
\small
\scalebox{0.95}{
\setlength{\tabcolsep}{4pt}
\begin{tabular}{lcccccc}
\hline
 & \multicolumn{2}{c}{R1@0.5} & \multicolumn{2}{c}{R1@0.7} & \multicolumn{2}{c}{mAP} \\
Model & sal. & dec. & sal. & dec. & sal. & dec. \\
\hline
TaskWeave (HD$\to$MR) & \textbf{51.9} & 37.6 & \textbf{36.5} & 19.2 & \textbf{29.9} & 17.0 \\
UVCOM & \textbf{50.5} & 47.6 & \textbf{35.9} & 28.2 & \textbf{30.2} & 24.2 \\
TR-DETR & 41.1 & \textbf{52.2} & 22.7 & \textbf{32.9} & 18.2 & \textbf{26.7} \\
\hline
\end{tabular}
}\vspace{-5mm}
\end{table}

\section{Experiments}
\label{sec:experiments}
\subsection{Setup}
\label{ssec:setup}
\textbf{Dataset.}
CASTELLA~\cite{munakata2025castella} contains 2,182 / 352 / 1,347 queries in its train / validation / test splits, with recordings of up to 300~s.
Test queries contain 3.1 moments on average; 57\% contain at least two moments and 27\% contain at least four.
Moment boundaries are annotated at 1~s resolution.

\para\textbf{Features.}
M2D-CLAP~\cite{niizumi2024m2dclap} encodes the audio into one 3840-dimensional vector per second and the query into 768-dimensional token features.
Both are projected to $d=256$ using two blocks consisting of a linear layer, layer normalization, and dropout, with a ReLU between the blocks.

\para\textbf{Models.}
We use QD-DETR (the architecture used in AM-DETR) and CG-DETR as the primary models, and additionally evaluate TaskWeave, UVCOM, and TR-DETR.
All models are implemented using Lighthouse~\cite{nishimura2024lighthouse}.
Each model produces $N_a=10$ decoder candidates, and we retain its original loss weights except in the training-objective experiments of Sec.~\ref{ssec:exp_training}.

\para\textbf{Training.}
We train each model for 200 epochs with a batch size of 32 using AdamW with a learning rate of $10^{-4}$, weight decay of $10^{-4}$, and gradient clipping at 0.1.

\para\textbf{Segmentation hyperparameters.}
For the segmentation in Sec.~\ref{ssec:segmentation}, we select one shared setting using validation R1 across configurations: $\alpha=0.5$, $g=2$ frames (2~s), and $\tau=0.7$.
These values are fixed across all models and experiments.
We searched $\alpha$ from 0.30 to 0.70 in steps of 0.05, $g$ over the integers from 0 to 6, and $\tau$ from 0.5 to 0.9 in steps of 0.1.

\para\textbf{Metrics.}
Before evaluation, both outputs are converted to seconds, clipped to the recording boundaries, and rounded to a 1-s grid; no minimum duration is enforced.
Recall@1 (R1) measures the fraction of queries whose top-ranked prediction overlaps any annotated moment at the specified IoU threshold.
We report R1@0.5 and R1@0.7. For mAP, we use at most ten predictions per query and average AP over queries and IoU thresholds from 0.50 to 0.95 in steps of 0.05~\cite{lei2021momentdetr}.
Results are test-set means over three seeds; SDs use divisor $n=3$, and differences are computed before rounding the seed means.

\subsection{Saliency- and decoder-based prediction}
\label{ssec:exp_main}
Unless otherwise stated, both outputs are evaluated at the same epoch selected by saliency-based validation R1@0.7 in each run.
Table~\ref{tab:main} compares the resulting paired predictions.
Surprisingly, simply segmenting the auxiliary saliency scores outperforms the dedicated moment decoder in all 18 paired QD-DETR and CG-DETR runs (6 configurations $\times$ 3 seeds), with an average R1@0.7 gain of 13.2 points.
Saliency-based prediction also achieves higher mean R1@0.5 and mAP in every configuration.

The mean R1@0.7 gap is 15--16 points for QD-DETR and 10--11 points for CG-DETR.
Across the three loss weights, CG-DETR's decoder-based score exceeds QD-DETR's by 4.9--7.3 points, whereas their saliency-based scores differ by only 0.6--2.2 points.
The smaller gap for CG-DETR therefore mainly reflects its higher decoder-based score rather than a lower saliency-based score.
We do not isolate which component of CG-DETR accounts for this difference.

To assess whether saliency-based epoch selection affects the comparison, we additionally select epochs by decoder-based validation R1@0.7 (Table~\ref{tab:epochs}).
This changes decoder-based test R1@0.7 by $-0.3$ to $+3.8$ points (2.2 points on average), but the resulting scores remain 7--15 points below the saliency-based scores in all six configurations.

\para\textbf{Effect of the segmentation rule.}
Table~\ref{tab:variants} tests whether the saliency-based advantage depends on the peak-relative segmentation rule in Sec.~\ref{ssec:segmentation} by applying simpler rules to the same frame-wise scores in Eq.~(\ref{eq:saliency}).
A recording-wise threshold already achieves 32.5--35.4 R1@0.7, compared with 21.0--28.4 for decoder-based prediction, showing that the advantage is not specific to our segmentation rule.
The peak-relative rule further improves R1@0.7 by 3.1--4.3 points.
Removing gap tolerance reduces R1@0.7 by 5.1--5.9 points; overlap removal, also disabled in this variant, preserves the top-ranked candidate and therefore does not affect R1.
In contrast, a single global absolute threshold reduces R1@0.7 to 21.2--25.1.
A fixed 5-s duration performs substantially worse (6.0--6.5), supporting the use of input-dependent durations.
\begin{table}[t]
\centering
\caption{Test performance (\%) by the number of annotated moments (mean over three seeds; $n$ = number of test queries). Both metrics use IoU 0.7; moment recall uses up to ten predictions and is averaged over queries. sal./dec.: saliency-/decoder-based predictions.}
\label{tab:strat}
\scalebox{0.95}{
\small
\begin{tabular}{llccc}
\hline
 & & 1 moment & 2--3 & $\geq$4 \\
Model & Output & ($n$=573) & ($n$=404) & ($n$=370) \\
\hline
\multicolumn{5}{l}{\textit{R1@0.7}} \\
QD-DETR ($\lambda=1$) & dec. & 21.8 & 21.3 & 19.3 \\
                      & sal. & 34.3 & 36.1 & 38.9 \\
CG-DETR ($\lambda=8$) & dec. & 29.7 & 26.7 & 28.5 \\
                      & sal. & 36.9 & 38.4 & 41.0 \\
\hline
\multicolumn{5}{l}{\textit{Moment recall@0.7 (top 10)}} \\
QD-DETR ($\lambda=1$) & dec. & 29.9 & 21.5 & 13.2 \\
                      & sal. & 50.8 & 38.6 & 30.4 \\
CG-DETR ($\lambda=8$) & dec. & 39.3 & 29.5 & 18.8 \\
                      & sal. & 51.2 & 37.6 & 28.5 \\
\hline
\end{tabular}
}\vspace{-4mm}
\end{table}

\textbf{Other DETR-based architectures.}
Table~\ref{tab:families} extends the comparison to TaskWeave in its highlight-detection-first configuration~\cite{yang2024taskweave}, UVCOM~\cite{xiao2024uvcom}, and TR-DETR~\cite{sun2024trdetr}, using the same segmentation rule without adding any components for saliency-based prediction.
Saliency-based prediction improves R1@0.7 by 17.3 points on TaskWeave and 7.7 points on UVCOM, with higher scores in all three seeds for both models.
TR-DETR shows the opposite behavior.
Its decoder-based prediction reaches 32.9, compared with 22.7 for saliency-based prediction.

Unlike the other models examined here, TR-DETR does not obtain its saliency from the encoder alone.
It uses the highest-scoring decoder moment to condition the frame representations used for saliency prediction~\cite{sun2024trdetr}.
Its saliency sequence therefore depends explicitly on the decoder's top prediction.
This contrast is consistent with a role for saliency construction, but the cross-model comparison does not isolate its effect.

\subsection{Performance gap by the number and duration of moments}
\label{ssec:exp_strat}
Table~\ref{tab:strat} reports R1@0.7 and moment recall by the number of annotated moments.
Moment recall is the matched fraction of annotations per query, averaged over queries, using maximum one-to-one matching with up to ten highest-ranked predictions at IoU $\geq 0.7$.
Saliency-based prediction is higher on both metrics in every group.
The R1@0.7 gap increases with moment count, whereas the moment-recall gap is smaller in the multi-moment groups than in the single-moment group.

Table~\ref{tab:length} reports R1@0.7 by the mean duration of the query's annotated moments.
For queries whose annotated moments have a mean duration of at most 2~s, saliency-based R1@0.7 reaches 27.8 and 31.6, compared with 6.3 and 12.3 for decoder-based prediction, and the gap narrows as the mean moment duration increases.
For each annotated moment lasting at most 2~s, we check whether any of the ten highest-ranked candidates reaches an IoU of at least 0.7.
Averaged over three seeds, this moment-level recall is 21.3\% for saliency-based prediction versus 3.2\% for decoder-based prediction in QD-DETR ($\lambda=1$), and 20.8\% versus 7.6\% in CG-DETR ($\lambda=8$).
Thus, the gap persists when candidates below the top rank are considered.
On the 1-s grid, an IoU of at least 0.7 with a 1- or 2-s moment requires both predicted endpoints to match the annotation exactly.

\begin{figure}[t]
\centering
\includegraphics[width=\linewidth]{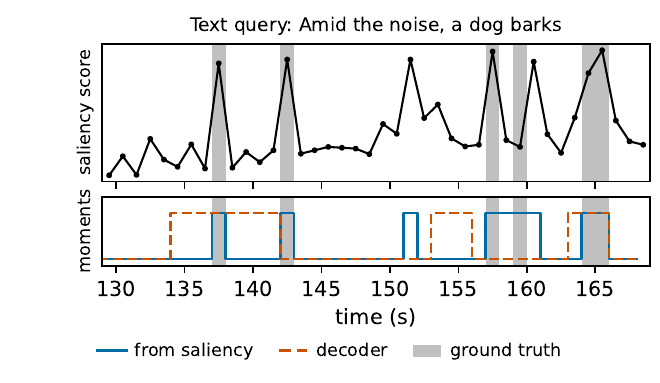}\vspace{-5mm}
\caption{
Example from CG-DETR ($\lambda=8$, seed 2023) on CASTELLA test, cropped to 129--168~s.
Top: raw saliency scores $r_j$ and annotated moments. Bottom: the five highest-ranked predictions of each output.
}
\label{fig:short}\vspace{-0mm}
\end{figure}

Figure~\ref{fig:short} illustrates these different prediction patterns for a recording containing several 1-s moments.
Saliency-based predictions more tightly capture several ground-truth moments, whereas the decoder produces longer segments.
Both outputs nevertheless miss or inaccurately delineate some annotated moments.

\begin{table}[t]
\centering
\caption{Test R1@0.7 (\%) by the mean duration of the query's annotated moments (mean over three seeds; $n$ = number of test queries). sal./dec.: saliency-/decoder-based predictions.}
\label{tab:length}
\small
\begin{tabular}{llccc}
\hline
 & & $\leq$2~s & $(2,5]$~s & $>$5~s \\
Model & Output & ($n$=290) & ($n$=358) & ($n$=699) \\
\hline
QD-DETR ($\lambda=1$) & dec. & 6.3 & 15.7 & 29.7 \\
                      & sal. & 27.8 & 30.7 & 42.3 \\
CG-DETR ($\lambda=8$) & dec. & 12.3 & 19.5 & 39.7 \\
                      & sal. & 31.6 & 30.7 & 45.4 \\
\hline
\end{tabular}\vspace{-4mm}
\end{table}

\subsection{Effect of the training objective}
\label{ssec:exp_training}
Increasing the saliency-loss weight $\lambda$ in Eq.~\eqref{eq:loss} from 1 to 8 raises mean saliency-based R1@0.7 by 1.1 points for QD-DETR and 1.8 points for CG-DETR as shown in Table~\ref{tab:main}.
Mean decoder-based R1@0.7 also increases, by 0.1 and 2.5 points, respectively.
The saliency-based advantage is already substantial under the original loss weighting ($\lambda=1$).

At $\lambda=8$, removing decoder supervision while retaining the other losses reduces mean saliency-based R1@0.7 by 2.9 points for QD-DETR and 2.2 points for CG-DETR (Table~\ref{tab:salonly}), with similar decreases in mAP.
These results suggest that decoder supervision remains useful during training even when the decoder's own predictions are less accurate than saliency-based predictions.

\begin{table}[t]
\centering
\caption{
Saliency-based prediction with joint training and without decoder supervision (no dec.), retaining all other losses ($\lambda=8$; test, \%, mean $\pm$ SD over three seeds).
}
\label{tab:salonly}
\small
\begin{tabular}{lcccc}
\hline
 & \multicolumn{2}{c}{R1@0.7} & \multicolumn{2}{c}{mAP} \\
Model & joint & no dec. & joint & no dec. \\
\hline
QD-DETR & 37.3$\pm$1.0 & 34.3$\pm$3.2 & 31.4$\pm$0.4 & 29.0$\pm$1.8 \\
CG-DETR & 38.5$\pm$0.6 & 36.4$\pm$1.9 & 31.4$\pm$0.1 & 29.4$\pm$1.4 \\
\hline
\end{tabular}\vspace{-0mm}
\end{table}

\section{Conclusion}
\label{sec:conclusion}
This study shows that auxiliary frame-wise saliency can provide substantially better moment predictions than decoder output in DETR-based AMR.
On CASTELLA, saliency-based prediction outperforms decoder-based prediction for QD-DETR and CG-DETR across all configurations, with R1@0.7 gaps of 10--16 points.
The R1@0.7 gap is larger for queries with more annotated moments and for those whose moments are shorter on average.
The advantage extends to TaskWeave and UVCOM but reverses for TR-DETR, whose saliency explicitly depends on the decoder prediction.
Decoder supervision also benefits saliency-based prediction.
Further work should examine how duration and temporal resolution affect saliency- and decoder-based retrieval across datasets, to understand their different behavior and improve both prediction pathways.

\bibliographystyle{IEEEbib}
\bibliography{strings,refs}

\end{document}